\documentclass[12pt,a4paper]{article}
\usepackage{amsmath,amssymb,booktabs,array,graphicx,tabulary}
\usepackage[numbers,sort]{natbib}
\setcitestyle{citesep={], [}}
\usepackage[a4paper,textwidth=6.8in,top=25mm,bottom=25mm,centering]{geometry}
\usepackage{flafter}
\usepackage[hidelinks]{hyperref}
\makeatletter
\renewcommand{\NAT@def@citea}{\def\@citea{\NAT@sep}}
\AtBeginDocument{%
  \if@filesw\immediate\write\@auxout{\string\citation{TPBibControl}}\fi
}
\makeatother
\ifdefined\pdfinfoomitdate\pdfinfoomitdate=1\relax\fi
\ifdefined\pdfsuppressptexinfo\pdfsuppressptexinfo=-1\relax\fi
\begin{document}
\begin{center}
{\Large\bfseries Three-body streaming forces lower the axial binding threshold of a held transverse three-sphere chain\par}
\vspace{0.7em}
Julian Olszewski\textsuperscript{1,\ensuremath{\dagger}},
Zijian Liu\textsuperscript{1,\ensuremath{\dagger}},
Erfan Eskandari\textsuperscript{2},
Yang Lin\textsuperscript{2},
Yuan Gao\textsuperscript{3}, and Jie Xu\textsuperscript{1,*}\\[0.5em]
{\small
\textsuperscript{1}Department of Mechanical and Industrial Engineering, University of Illinois Chicago,\\
842 W. Taylor Street, MC 251, Chicago, IL 60607, USA\\
\textsuperscript{2}Department of Mechanical, Industrial and Systems Engineering, University of Rhode Island, Kingston, RI 02881, USA\\
\textsuperscript{3}Department of Mechanical Engineering, University of Memphis, Memphis, TN 38152, USA}
\end{center}
\noindent{\small \textsuperscript{\ensuremath{\dagger}}\,Julian Olszewski and Zijian Liu contributed equally to this work.}\par
\noindent{\small \textsuperscript{*}Corresponding author: \href{mailto:jiexu@uic.edu}{jiexu@uic.edu}.}\par
\noindent{\footnotesize ORCIDs: J.~Olszewski, \href{https://orcid.org/0009-0006-2346-8097}{0009-0006-2346-8097};
Z.~Liu, \href{https://orcid.org/0009-0000-6014-1268}{0009-0000-6014-1268};
E.~Eskandari, \mbox{\href{https://orcid.org/0000-0002-4934-7609}{0000-0002-4934-7609}};\\
Y.~Lin, \mbox{\href{https://orcid.org/0000-0002-8261-8984}{0000-0002-8261-8984}};
Y.~Gao, \href{https://orcid.org/0000-0002-5464-1626}{0000-0002-5464-1626};
J.~Xu, \href{https://orcid.org/0000-0001-5765-7431}{0000-0001-5765-7431}.}\par
\medskip
\noindent\textbf{Abstract}\par\smallskip

\noindent Particles in oscillatory liquids can form chains perpendicular to the drive. We ask whether the forces between isolated pairs predict when three spheres can bind at a finite spacing. We first hold the spheres fixed during small-amplitude oscillation and calculate the cycle-averaged force on an end sphere. The three-body contribution lowers the critical Stokes number from 4.283 for the pair-additive model to 3.534 for the full chain, a reduction of 17.5 percent. An independent finite-element calculation reproduces the shift. The three-body attraction also draws the spheres close enough for the nearest-neighbour pair forces to become repulsive. This compression accounts for most of the tendency to bend out of the plane containing the drive and the chain. For freely translating and rotating spheres at a particle-to-liquid density ratio of 2.2, the multipole calculation predicts initial growth of this bend at five sampled equilibria. A point-singularity calculation confirms the sign at two of those spacings.

\smallskip\par\noindent\textit{Short title: Three-body streaming and axial binding}
\medskip

\section{Introduction}

Particles in an oscillatory liquid can form rows across the drive \citep{Wunenburger2002,Klotsa2009,vanOverveld2023}. A transverse pair can settle at a finite separation when its time-averaged force changes from repulsion at short distances to attraction at larger distances \citep{Fabre2017,ZhangRallabandi2025}. This gives a natural starting point for a model of a chain. Effective pair potentials have, for example, been used to model the self-organisation of particles in oscillating flows \citep{vanOverveld2024Siren}. In a three-sphere chain, each end sphere interacts with two neighbours at different distances. Adding those two isolated-pair forces predicts a net force and a binding threshold.

\citet{Kyotoh1991} predicted that three equal spheres aligned across an oscillatory flow would approach one another in the limit of large separations and small sphere Reynolds number. The existence of a restoring spacing depends on how the mean axial force changes as the spheres draw closer. Near the binding threshold, the maximum of the pair-additive force is close to zero. An attractive three-body contribution can lift that maximum above zero, opening an interval of attraction with a restoring spacing at its inner edge. We ask whether this occurs at Stokes numbers for which the pair-additive chain has no such spacing.

In a separate U-tube observation, \citet{Kyotoh1991} reported that a transverse row of three spheres rolling on the floor initially moved regularly, then split into a pair and a single sphere. In air, \citet{Wu2025ThreeBody} computed a slightly larger stable separation in an equilateral acoustic trimer than in a pair. They also quantified force nonadditivity in a collinear trimer by subtracting isolated-pair forces. \citet{MaoJaeger2026} showed that non-conservative three-body acoustic forces can select a stable flattened triangle. Corrections to pair superposition also enter the Brownian viscosity of hard spheres \citep{OlszewskiLiuXu2026}. Here the issue is how the three-body streaming force in a liquid changes the onset of binding and the shape response of a straight chain.

We compare the full three-sphere force with the sum of the two isolated-pair forces, holding the spheres fixed during the drive in both calculations. We locate the axial fold, where the stable and unstable axial equilibrium spacings merge. This gives the critical Stokes number for the finite-spacing branch. We then test how the spacing selected by the full force responds to axial and bending perturbations. A separate calculation allows the spheres to translate and rotate during the drive and checks whether the out-of-plane bend still grows.

\section{Problem and method}

\subsection{Geometry and force laws}

\begin{figure}[!htbp]
\centering
\includegraphics[width=\textwidth]{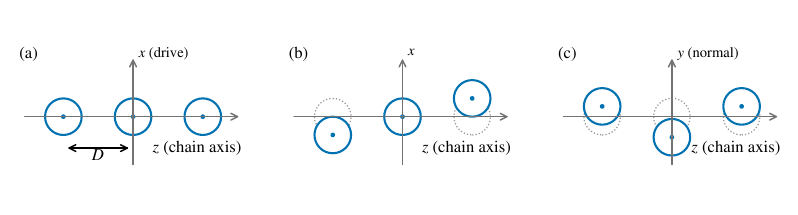}
\caption{Straight chain and transverse shape modes. (a) The $xz$ view shows the chain along $z$, the drive along $x$, and the centre spacing $D$. (b) Tilt towards the drive in the same view. (c) Normal bend in the $yz$ view. The in-plane bend uses the pattern in (c) along $x$. Dotted circles mark the straight reference configuration.}
\label{fig:geometry}
\end{figure}

Identical rigid spheres of radius $a$ lie on the $z$-axis with dimensionless centre-to-centre spacing $D$ between neighbours. The liquid has density $\rho$ and kinematic viscosity $\nu$. Far from the spheres, its velocity is $U\cos(\omega t)\mathbf e_x$, perpendicular to the chain (Figure~\ref{fig:geometry}). The Stokes number is $\Omega=\omega a^2/\nu$. We scale length by $a$, velocity by $U$, and time by $1/\omega$. The dimensionless excursion $\epsilon=U/(\omega a)$ is small compared with unity and with the nearest surface gap $D-2$. The streaming force is the mean hydrodynamic force at order $U^2$, scaled by $\rho U^2a^2$.

We use the cycle-averaged axial force on an end sphere, taking attraction towards the chain interior as positive. For the full held chain this force is $F_3(D)$. The isolated-pair forces $F_2(D)$ and $F_2(2D)$ are calculated separately for the nearer and farther neighbour, at dimensionless spacings $D$ and $2D$. Their sum is $F_{\rm add}(D)=F_2(D)+F_2(2D)$. The connected force is
\begin{equation}
\Delta F=F_3(D)-F_{\rm add}(D).
\label{eq:connected}
\end{equation}
The pair sum already includes the direct force from the farther sphere, and an isolated sphere has zero mean force by symmetry. Thus $\Delta F$ isolates the part of the end force that depends on all three spheres being present. We recompute the pair forces with the same flow model and boundary conditions used for the full chain. This also supplies the spacing derivatives at both pair separations needed to locate the additive fold.

\subsection{Flow and particle motion}

We expand the dimensionless velocity as $\mathbf u=\mathbf u_1+\epsilon\mathbf u_2+\cdots$. For the oscillatory velocity $\operatorname{Re}(\mathbf v e^{-\mathrm i t})$ and the mean second-order velocity $\overline{\mathbf u}_2$, the flow equations are
\begin{equation}
\begin{aligned}
-\mathrm i\mathbf v &= -\nabla p+\Omega^{-1}\nabla^2\mathbf v, & \nabla\cdot\mathbf v &=0,\\
\Omega^{-1}\nabla^2\overline{\mathbf u}_2-\nabla\bar p_2 &= \mathbf b, & \nabla\cdot\overline{\mathbf u}_2 &=0,
\end{aligned}
\label{eq:flow}
\end{equation}
where $\mathbf b=\tfrac12\operatorname{Re}[(\overline{\mathbf v}\cdot\nabla)\mathbf v]$ and the overbar on a complex amplitude denotes conjugation. At infinity $\mathbf v\to\mathbf e_x$ and $\overline{\mathbf u}_2\to0$. The held calculation fixes the positions and orientations of the spheres during the drive, so both velocities vanish on their surfaces. We evaluate this held force law at nearby configurations and apply steady mobility to infer the response to slow changes of shape.

On a free sphere $k$, no slip at first order gives $\mathbf v=\mathbf V_k+\boldsymbol\Omega_k\times(\mathbf x-\mathbf x_{k0})$ on its reference surface. The translation and angular velocity amplitudes follow from
\begin{equation}
-\mathrm i\,\tilde\rho\,\frac{4\pi}{3}\mathbf V_k=\mathbf F^{(1)}_k,\qquad
-\mathrm i\,\tilde\rho\,\frac{8\pi}{15}\boldsymbol\Omega_k=\mathbf T^{(1)}_k,
\label{eq:rigid_motion}
\end{equation}
where $\tilde\rho=\rho_p/\rho$ is the particle-to-liquid density ratio and $\mathbf F^{(1)}_k$ and $\mathbf T^{(1)}_k$ are the first-order fluid force and torque. The force includes the imposed pressure-gradient contribution $-\mathrm i(4\pi/3)\mathbf e_x$. The displacement amplitude is $\boldsymbol\xi_k=\mathrm i\mathbf V_k$. We neglect gravity and other body forces in the particle dynamics. The free-chain results in the main text use $\tilde\rho=2.2$.

The reciprocal theorem gives each held force component as $-\int_V\hat{\mathbf u}\cdot\mathbf b\,dV$, where $V$ is the liquid volume and $\hat{\mathbf u}$ is the steady Stokes flow for unit translation of the target sphere with the others fixed \citep{Fabre2017,ZhangRallabandi2025}. For free spheres we evaluate the mean loads in the translating frame of the target $j$. In that frame the advecting velocity in the quadratic forcing is $\mathbf v-\mathbf V_j$. We apply the moving-wall expansion of the no-slip condition on the mean surface \citep[App.~A, Eq.~(A6)]{Liu2026CavityStreaming}. For sphere $k$, the displacement in that expansion is $\boldsymbol\xi_k-\boldsymbol\xi_j$. A unit rotation of the target gives the auxiliary flow for a torque component. We set the mean rigid velocities to zero to obtain the streaming loads, then apply steady mobility to calculate the slow drift. Supplementary Section S1 gives the stress, mean boundary condition and full reciprocal identity.

\subsection{Numerical discretisation}

The free-space solver expands the first-order disturbance and the steady auxiliary flow in vector spherical multipoles centred on each sphere. We impose the surface conditions by least-squares collocation and check their residuals on a rotated set of points. The reciprocal volume integral is split by planes normal to the chain axis, halfway between the axial coordinates of neighbouring centres. The reference degree is $L=14$, increasing to 20 for $2.35\le D<2.7$ and to 28 for $D<2.35$. We check the held folds at $L=18$. For the small displaced configurations used to form the transverse derivatives, we retain azimuthal orders $|m|\le2$ about each sphere and check the result at 3.

A separate finite-element solver checks the held axial forces and folds. It uses quadratic velocity and linear pressure elements on a graded meridional mesh with a typical surface element size of $0.03a$ (Supplementary Figure~S1). The artificial outer boundary is at radial distance $60a$ and axial positions $\pm60a$. There we impose zero first-order disturbance and zero velocity in the mean and auxiliary problems. Reflection about $z=0$ allows a half-domain calculation. The axial auxiliary field assigns opposite velocities to the two end spheres. Both halves contribute equally to the reciprocal integral, so the half-domain result gives the force on one end. The transverse drive excites azimuthal orders $m=\pm1$ at first order. The quadratic forcing has orders $m=0,\pm2$, of which only $m=0$ contributes to the axial force because the axial auxiliary flow is axisymmetric. A time-domain calculation checks force increments at selected finite amplitudes. It integrates azimuthal orders 0 to 2 with second-order backward differences and explicit convection (Supplementary Section S2).

\subsection{Equilibria and shape response}

At the critical Stokes number $\Omega_c$, the force curve $F(D,\Omega_c)$ touches zero at $D_c$. We solve $F=\partial_D F=0$ on local $5\times3$ multipole force stencils using fits of degree four in $D$ and two in $\Omega$. We recenter the stencil at each new estimate. The finite-element check interpolates force maxima between two Stokes numbers that bracket the fold.

At each stable axial zero $D^*$, we form centred differences using multipole forces for shifts of $\pm0.02a$ along $x$ or $y$. Reflection symmetry decouples these directions. The shape modes are tilt $(-1,0,1)$ and bend $(1,-2,1)$. For held spheres we use isolated-sphere drag and check the normal bend with Rotne--Prager and full Stokes mobility \citep{RotnePrager1969}.

For free spheres, let $K_F$ and $K_T$ be derivatives of the dimensionless mean loads with respect to transverse displacements in units of $a$. The force and torque scales are $\rho U^2a^2$ and $\rho U^2a^3$. With the steady mobility normalised by isolated-sphere drag as in Supplementary Section S4, the shape drift matrix is
\begin{equation}
P(\widehat{\mathsf M}_{UF}K_F+\widehat{\mathsf M}_{UT}K_T)P,\qquad
P=I-\tfrac13\mathbf1\mathbf1^{\mathsf T}.
\label{eq:shape_drift}
\end{equation}
The two blocks map forces and torques to translation in the perturbed direction. The mean loads vanish at the straight equilibrium, leaving only their derivatives in the linearisation. The projection $P$ removes uniform translation. Rotation about the drive makes $y$ tilt neutral. A dimensionless shape eigenvalue $\lambda$ gives a rate $\sigma=\lambda\rho U^2/(6\pi\mu)$, with dynamic viscosity $\mu=\rho\nu$.

\section{Connected force and axial binding}

\begin{figure}[!htb]
\centering
\includegraphics[width=0.96\textwidth]{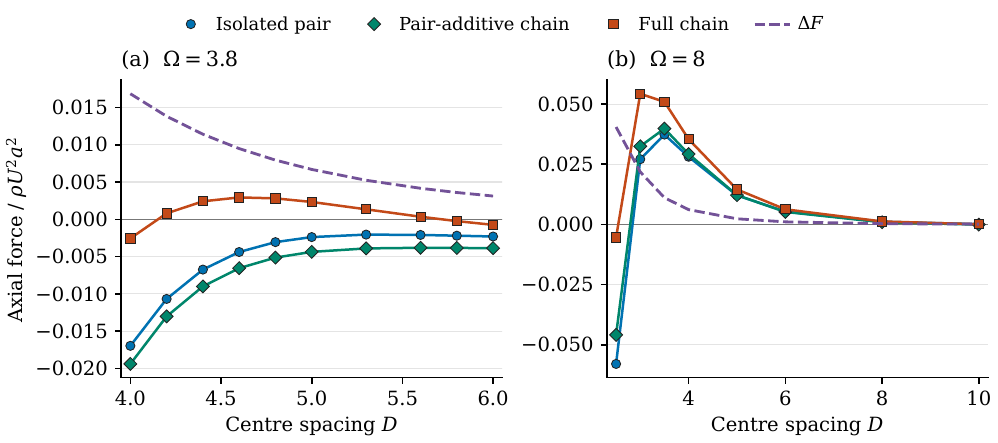}
\caption{Finite-element mean axial force as a function of spacing at (a) $\Omega=3.8$ and (b) $\Omega=8$. Markers show the computed forces. The purple dashed curve joins the values of $\Delta F=F_3-F_{\rm add}$ at the same spacings. Positive force denotes attraction.}
\label{fig:forces}
\end{figure}

The three-body force can change whether a chain has an equilibrium. At $\Omega=3.8$, the full force crosses zero on the inner side of its attractive interval, giving a restoring axial spacing. The pair-additive force remains repulsive over the same range of spacings (Figure~\ref{fig:forces}(a)). At the higher Stokes number $\Omega=8$, both force laws admit a restoring spacing. At $D=3$, the connected force adds about two thirds of the pair-additive force, and the full chain reaches equilibrium at a smaller spacing (Figure~\ref{fig:forces}(b)).

\begin{figure}[!htb]
\centering
\includegraphics[width=0.96\textwidth]{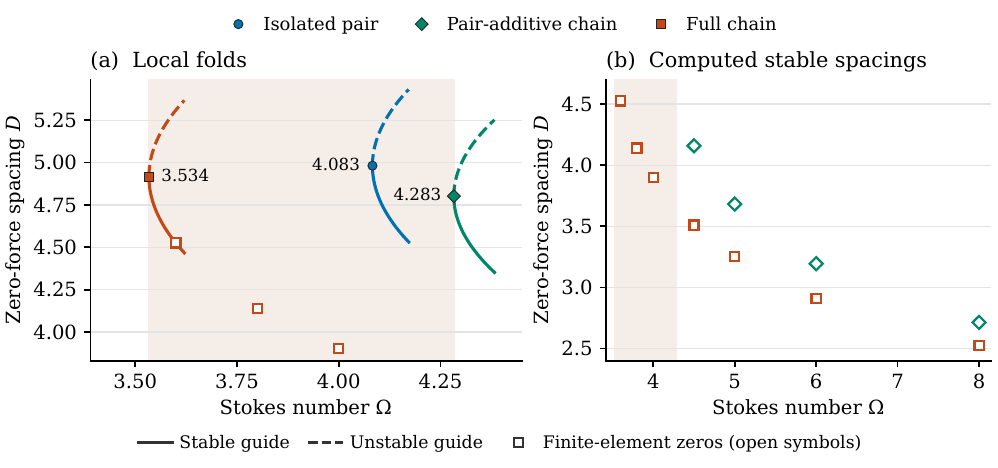}
\caption{Axial equilibrium spacings for held spheres. (a) Filled symbols mark the free-space multipole folds. Solid and dashed curves give the local stable and unstable branches from the quadratic expansion over $|D-D_c|\leq0.45$. Open symbols mark stable finite-element spacings of the full and pair-additive chains, with the full sampled range in (b). The shaded interval lies between their multipole folds.}
\label{fig:fold}
\end{figure}

The difference in binding can be traced to the two zero-force branches. On the inner branch, an increase in spacing increases attraction, which draws the end spheres back towards equilibrium. On the outer branch, the slope reverses and a displacement grows. At the fold $(D_c,\Omega_c)$, the branches meet and the axial restoring stiffness vanishes. Locally the force is $F\simeq A(\Omega-\Omega_c)-B(D-D_c)^2$, where $A=\partial_\Omega F$ and $B=-\tfrac12\partial_D^2 F$ at the fold. Both coefficients are positive in the three fitted force laws. Figure~\ref{fig:fold} shows the local branches and the computed stable equilibria. Supplementary Section S2 gives the fold coordinates and coefficients.

\begin{table}[!htbp]
\centering
\begin{tabular}{lrrr}
\toprule
Force law & Multipole $\Omega_c$ & Finite-element $\Omega_c$ & $\Delta\Omega_c$ \\
\midrule
Isolated pair & 4.0830 & 4.0828 & $-0.0002$ \\
Pair-additive chain & 4.2829 & 4.2806 & $-0.0023$ \\
Full held chain & 3.5338 & 3.5345 & $+0.0007$ \\
\bottomrule
\end{tabular}
\caption{Binding thresholds of the three held-sphere force laws. The finite-element domain extends to 60 sphere radii. Differences are finite element minus multipole, calculated before rounding. Supplementary Section S2 gives the fold spacings and sensitivity checks.}
\label{tab:fold}
\end{table}

The isolated-pair fold provides a reference for the two chain predictions. The farther pair is repulsive near the fold, so adding it raises the threshold above that for the isolated pair. The connected attraction then lowers the free-space multipole threshold from 4.283 to 3.534, a 17.5\% reduction relative to pairwise addition (Figure~\ref{fig:fold}). This opens a range of Stokes numbers in which the full model predicts axial binding. The finite-element shift is 0.746 (17.4\%), compared with 0.749 from multipoles (Table~\ref{tab:fold}). The largest solver difference is about 0.0023, comparable to the peak-fit sensitivity measured for the pair and full chain. Supplementary Section S2 gives separate mesh, domain, fit and multipole-order checks, and compares the additive fold with a point-singularity calculation.

The threshold shift compares force laws evaluated under matched conditions. Published pair forces provide separate checks on $F_2$. Reconstructing the outer boundary of \citet{Fabre2017} reduces the discrepancy in zero-force frequency at fixed pair spacing. The origin of the remaining difference is undetermined. The curves of \citet{ZhangRallabandi2025} provide a force comparison at selected spacings and frequencies. A direct surface-traction calculation on the reconstructed pair mesh also checks the reciprocal force evaluation. Supplementary Sections S2 and S3 document these comparisons and the time-domain consistency checks.

\section{Stability of the straight chain}

An axial equilibrium can still be unstable to other changes in shape. We first examine perturbations of the two gaps. In breathing they change together, and in dimerisation one opens while the other closes. Both motions are restoring at the sampled held equilibria (Supplementary Section S2). For disturbances transverse to the chain, tilt and bending along the drive are restoring with isolated-sphere drag, whereas bending normal to the drive--chain plane grows (Table~\ref{tab:modes}). The normal bend also grows at all five sampled equilibria with Rotne--Prager translational mobility and with the full steady Stokes mobility including mean torque (Supplementary Section S4).

\begin{table}[!htbp]
\centering
\begin{tabular}{lrrrrr}
\toprule
Held chain & $\Omega$ & $D^*$ & $x$ tilt & $x$ bend & $y$ bend \\
\midrule
Held & 3.6 & 4.525 & $-0.0118$ & $-0.0306$ & $+0.0047$ \\
Held & 5 & 3.251 & $-0.0991$ & $-0.272$ & $+0.0256$ \\
Held & 8 & 2.525 & $-0.320$ & $-0.881$ & $+0.0745$ \\
\bottomrule
\end{tabular}
\caption{Representative held-chain force gradients at the zero-force spacing $D^*$, in units of $\rho U^2a$. Negative gradients restore the shape with scalar mobility.}
\label{tab:modes}
\end{table}

The connected attraction draws the spheres close enough for the nearest-neighbour pair forces to be repulsive. When the middle sphere moves out of the plane, the repulsive bonds push it farther in that direction. For the bend $(1,-2,1)$ the two end spheres move together, so their direct pair interaction contributes no linear bending force. The pair contribution at the full-chain spacing $D^*$ is $-3F_2(D^*)/D^*$, and the full gradient minus this contribution gives the connected gradient at the same spacing (Supplementary Figure~S2).

At $\Omega=8$, the additive chain already has a weak positive gradient at its own equilibrium. Evaluating its force law at the shorter full-chain spacing increases that gradient by about 0.0515. The connected force adds about 0.0137 at that spacing, giving a total increase of 0.0652 in units of $\rho U^2a$.

Allowing translation and rotation during the drive changes the streaming loads and the equilibrium spacing. With mean torque included, the multipole normal-bend eigenvalue is positive at all five sampled equilibria. It is $2.18\times10^{-4}$ at $\Omega=3.6$ and $3.79\times10^{-3}$ at $\Omega=8$. The point-singularity values at the same multipole spacings are $2.17\times10^{-4}$ and $3.77\times10^{-3}$. Supplementary Section S4 reports the axial residuals at those spacings, the numerical sensitivity and the full set of rates.

\section{Discussion}

The full held chain can restore its axial spacing while amplifying a bend out of the plane. A planar constraint would suppress that mode. A physical wall would also change the forces and the mobility, and could therefore change the axial spacing itself.

The threshold applies to spheres held fixed during a small-amplitude drive in an unbounded liquid. The time-domain runs give a limited check on selected force increments. The individual period means have residual drift, and the finite-amplitude correction to the fold remains open. The U-tube observation reported by \citet{Kyotoh1991} used a floor and a liquid excursion about 31 sphere radii. The unequal-gap breakup observed there motivates testing how the axial modes respond to confinement and finite excursion. Experiments in horizontally oscillated cells show transverse rows near a floor \citep{Wunenburger2002,Klotsa2009,vanOverveld2023}. Numerical studies show how the floor, the Stokes layer and particle rotation affect the streaming flow and the selected spacing \citep{Mazzuoli2016,vanOverveld2021,vanOverveld2022Wall}.

An experimental comparison can use the equilibrium spacing as the observable. Pair experiments have resolved this gap, and simulations of the same system have determined the interaction force \citep{Klotsa2007}. For a trimer in the same cell, calculate the two pair forces over a range of spacings under matched drive, particle motion and confinement. Their sum at $D$ and $2D$ predicts whether a restoring spacing exists. The presence of a straight trimer and its measured spacing then test the pair-based prediction. A force baseline inferred from slow pair motion would additionally require the pair mobility, including wall and rotational effects, to be calibrated.

A separate test of the free-sphere calculation would perturb the middle sphere normal to the drive--chain plane and track its initial cycle-averaged offset relative to the outer spheres. The calculation neglects gravity, so a test away from walls would need to control sedimentation and quantify the effect of any supporting field on the bend. Magnetic levitation has been used for two spheres in a vibrated liquid \citep{PachecoMartinez2013}. The field provides confinement as well as support, so its contribution to the bending response would need to be calibrated. For spheres of radius $0.5$ mm in water with $\nu=10^{-6}\,\mathrm{m}^2\,\mathrm{s}^{-1}$, $\Omega=8$ corresponds to $5.1$ Hz. At density ratio $2.2$ and excursion $\epsilon=0.03$ ($15\,\mu\mathrm m$), the leading-order free-chain calculation gives an initial linear e-folding time of about $1.1\times10^5$ drive periods, or $6$ h. The growth time scales as $\epsilon^{-2}$ within the small-excursion regime (Supplementary Section S4).

\begin{samepage}
\section{Conclusion}

The three-body streaming force lowers the axial binding threshold of a transverse three-sphere chain held fixed during small-amplitude oscillation. The critical Stokes number falls from 4.283 for the pair-additive model to 3.534 for the full chain, a reduction of 17.5\%. Between these thresholds, only the full force law admits a restoring axial spacing on the branch studied here.

Three-body attraction draws the spheres close enough for the nearest-neighbour pair forces to be repulsive. If the middle sphere moves out of the plane containing the drive and the chain, those repulsive forces gain components that push it farther out. This compression accounts for most of the normal-bend force gradient at the sampled held equilibria. For freely translating and rotating spheres, the multipole calculation also predicts initial growth of this bend at all five sampled equilibria. A point-singularity calculation confirms the sign at two of those spacings.

\end{samepage}

\section*{Declarations}

\subsection*{Acknowledgments}
Claude Opus 5.5 and Fable 5.2 (Anthropic, through Claude Code) and GPT-6.1 (OpenAI, through Codex) assisted with literature verification, code development, numerical execution, data analysis and provided editorial assistance with the manuscript under the authors' direction.  Deterministic numerical solvers produced the results, and plotting scripts produced the figures. The authors are responsible for the accuracy, originality and final content of this work.

\subsection*{Informed consent}
Not applicable.

\subsection*{Author contributions}
All authors have accepted responsibility for the entire content of this manuscript and approved its submission.\par
  \textbf{Julian Olszewski:} Conceptualization, Methodology, Software, Formal analysis, Investigation,
  Funding acquisition, Writing -- original draft, Writing -- review and editing.\par
  \textbf{Zijian Liu:} Conceptualization, Methodology, Software, Formal analysis, Investigation,
  Writing -- original draft, Writing -- review and editing.\par
  \textbf{Erfan Eskandari:} Writing -- review and editing.\par
  \textbf{Yang Lin:} Funding acquisition, Writing -- review and editing.\par
  \textbf{Yuan Gao:} Writing -- review and editing.\par
  \textbf{Jie Xu:} Conceptualization, Investigation, Supervision, Writing -- review and editing.

\subsection*{Conflict of interest}
The authors state no conflict of interest.

\subsection*{Research funding}
J.~Olszewski acknowledges support from the University of Illinois Chicago
  Chancellor's Undergraduate Research Award (CURA).
Y.~Lin acknowledges support from the U.S. National Science Foundation under CAREER Award No.~2540390.

\subsection*{Data availability}
The full data and code supporting the findings of this study are not available before acceptance. They will be deposited in INDIGO, the University of Illinois Chicago's institutional repository, upon acceptance.

\subsection*{Research ethics}
This numerical study did not require human or animal ethics approval.

\bibliographystyle{tp_references}
\bibliography{references}
\end{document}


\begin{center}
{\Large\bfseries Supplementary material for Three-body streaming forces lower the axial binding threshold of a held transverse three-sphere chain\par}
\vspace{0.7em}
Julian Olszewski\textsuperscript{1,\ensuremath{\dagger}},
Zijian Liu\textsuperscript{1,\ensuremath{\dagger}},
Erfan Eskandari\textsuperscript{2},
Yang Lin\textsuperscript{2},
Yuan Gao\textsuperscript{3}, and Jie Xu\textsuperscript{1,*}\\[0.5em]
{\small
\textsuperscript{1}Department of Mechanical and Industrial Engineering, University of Illinois Chicago,\\
842 W. Taylor Street, MC 251, Chicago, IL 60607, USA\\
\textsuperscript{2}Department of Mechanical, Industrial and Systems Engineering, University of Rhode Island, Kingston, RI 02881, USA\\
\textsuperscript{3}Department of Mechanical Engineering, University of Memphis, Memphis, TN 38152, USA}
\end{center}
\noindent{\small \textsuperscript{\ensuremath{\dagger}}\,Julian Olszewski and Zijian Liu contributed equally to this work.}\par
\noindent{\small \textsuperscript{*}Corresponding author: \href{mailto:jiexu@uic.edu}{jiexu@uic.edu}.}\par
\noindent{\footnotesize ORCIDs: J.~Olszewski, \href{https://orcid.org/0009-0006-2346-8097}{0009-0006-2346-8097};
Z.~Liu, \href{https://orcid.org/0009-0000-6014-1268}{0009-0000-6014-1268};
E.~Eskandari, \mbox{\href{https://orcid.org/0000-0002-4934-7609}{0000-0002-4934-7609}};\\
Y.~Lin, \mbox{\href{https://orcid.org/0000-0002-8261-8984}{0000-0002-8261-8984}};
Y.~Gao, \href{https://orcid.org/0000-0002-5464-1626}{0000-0002-5464-1626};
J.~Xu, \href{https://orcid.org/0000-0001-5765-7431}{0000-0001-5765-7431}.}\par
\medskip

\section*{S1. Governing equations and force}

With the scales defined in the main text, the dimensionless equations are $\partial_t\mathbf u+\epsilon\,\mathbf u\cdot\nabla\mathbf u=-\nabla p+\Omega^{-1}\nabla^2\mathbf u$ and $\nabla\cdot\mathbf u=0$. The velocity expansion leading to Eq.~(2) is accompanied by $\mathbf x_k=\mathbf x_{k0}+\epsilon\mathrm{Re}(\boldsymbol\xi_k e^{-\mathrm{i}t})+\dots$ for the sphere centres. First-order pressure is scaled by $\rho U\omega a$ and mean second-order stress by $\rho U^2$.

\emph{First-order traction.} The ambient pressure amplitude is $\mathrm{i}\,\mathbf e_x\cdot\mathbf x$. The normal $\mathbf n$ points from the solid into the liquid. The hydrodynamic force in Eq.~(3) of the main text is $\int_{S_k}\boldsymbol\sigma_1\mathbf n\,dS$ with $\boldsymbol\sigma_1=-p\mathsf I+\Omega^{-1}(\nabla\mathbf v+\nabla\mathbf v^{\rm T})$. The torque uses the moment of the same traction about $\mathbf x_{k0}$.

\emph{Mean second order.} The second row of Eq.~(2) in the main text uses $\bar{\mathbf u}_2=0$ on held spheres. On a translating sphere the condition is transferred to the mean surface,
$\bar{\mathbf u}_2=\bar{\mathbf V}^{(2)}_k+\bar{\boldsymbol\Omega}^{(2)}_k\times\mathbf r-\tfrac12\mathrm{Re}[(\bar{\boldsymbol
\xi}_k\cdot\nabla)\mathbf v]$. Pure rotation leaves a sphere's surface in place, so a rotating sphere needs the
Eulerian surface velocity rather than a translational shift term.

\emph{Force by the reciprocal theorem.} Let $(\hat{\mathbf u},\hat{\boldsymbol\sigma})$ be the Stokes flow of unit
viscosity in which the target sphere $j$ translates with unit velocity $\mathbf e'$, every other sphere is held and the
flow decays at infinity. Then
\begin{equation}
\mathbf e'\cdot\bar{\mathbf F}_j=-\int_V\hat{\mathbf u}\cdot\mathbf b\,dV+\Omega^{-1}\sum_k\int_{S_k}\bar{\mathbf u}_2\cdot
\hat{\boldsymbol\sigma}\mathbf n\,dS+\mathbf e'\cdot\int_{S_j}\tfrac12\mathrm{Re}[(\bar{\boldsymbol\xi}_j\cdot\nabla)
\boldsymbol\sigma_1]\mathbf n\,dS.
\end{equation}
For held spheres only the volume term remains. The freely oscillating calculations use zero gravity. We set the mean rigid velocities to zero when extracting the streaming loads. At slow time the free spheres move so that the steady Stokes resistance balances these loads. A cluster with a nonzero net force has a mean velocity decaying as $1/r$. The auxiliary stress decays as $1/r^2$, so the surface products at infinity still vanish. In a bounded
domain the outer surface term must be kept unless the primal and auxiliary conditions there make it vanish; with
$\bar{\mathbf u}_2=0$ and $\hat{\mathbf u}=0$ on the outer boundary of the finite-element domain it vanishes.

\emph{Symmetry of torques and rotation.} Spheres on the $z$-axis, drive along $x$. The mirrors
$M_x$ ($x\to-x$, which maps the drive to minus itself), $M_y$ and, for the symmetric chain, $M_z$ map the problem to
itself; a torque is a pseudovector and transforms under a mirror $R$ as $\det(R)\,R\mathbf T$. The mean force and torque
are quadratic in the drive, so the sign of the drive drops out: $M_x$ gives $(T_x,-T_y,-T_z)=\mathbf T$, so
$T_y=T_z=0$, and $M_y$ gives $T_x=0$. A displacement along $z$ keeps both mirrors, so the mean torque vanishes
identically for the straight chain and for every longitudinal deformation, by this symmetry. The first-order torque is
linear in the drive: $M_x$ gives $-\mathbf T=(T_x,-T_y,-T_z)$, so $T_x=0$, and $M_y$ gives $T_z=0$; only $T_y$, about
the axis normal to the drive and the chain, survives, and $M_z$ makes it vanish on the middle sphere of the symmetric
chain. The outer spheres of a free chain therefore rock about $y$ at first order. A displacement along $x$ breaks $M_x$
but keeps $M_y$, so it creates a mean torque $T_y$ alone at first order in the displacement; one along $y$ breaks
$M_y$ but keeps $M_x$ and creates $T_x$ alone. The computed derivatives obey these symmetry relations. $\partial T_y/\partial x_j$ reaches 0.05
and $\partial T_x/\partial y_j$ 0.0065 per radius at $\Omega=8$, and every other torque derivative is below $10^{-12}$.
For free spheres these torques drive the steady rotation whose effect on the transverse modes is computed in Section S4.

\emph{Amplitude ordering.} The second-order mean force is the leading term. In a finite-amplitude time-domain
calculation, third-order oscillatory fields multiply first-order ones in the fourth-order mean force, so the relative
correction is of order $\epsilon^2$ away from zeros of the leading force. With the force normalised by $\rho U^2a^2$,
the correction is of order $\epsilon^2$ in absolute terms, including near the folds. The time-domain runs used
azimuthal modes 0 to 2 at finite amplitude.

\emph{Connected forces.} For target $i$, $\mathbf F^{[3]}_i(i,j,k)=\mathbf F_i(i,j,k)-\mathbf F_i(i,j)-\mathbf
F_i(i,k)+\mathbf F_i(i)$, each term with the same held or free rule. For a held sphere the isolated mean force vanishes by symmetry.

\emph{Finite elements.} The gmsh target element size at distance $d$ from the nearest sphere is $h(d)/a=\min(2,0.03+0.12d/a)$ in the default mesh. The weak problems are assembled with scikit-fem. The symmetry plane $z=0$ has $u_z=0$ and zero tangential traction. For the axial auxiliary field in a symmetric layout, the mirrored end sphere translates with the opposite axial velocity. The half-domain integral then gives the force on a single end, since the two end forces in the full domain are equal and opposite.

\emph{Multipole basis.} The first-order disturbance is a sum over sphere centres of terms of the form $\nabla\Phi+\nabla\times(\mathbf x\chi)+\nabla\times\nabla\times(\mathbf x\psi)$, with $\mathbf x$ measured from the expansion centre. Here $\Phi$ is harmonic and $(\nabla^2-\lambda^2)(\chi,\psi)=0$ with $\lambda^2=-\mathrm{i}\Omega$. The auxiliary Stokes field uses Lamb's solution. The collocation grid uses Gauss--Legendre nodes in polar cosine and uniform azimuth. In each slab, sphere-centred volume quadrature uses Gauss--Legendre nodes in polar cosine and $\tau$, with radius $r=1+1.5\tau/(1-\tau)$ and $0<\tau<1$, and uniform azimuth. The reference off-axis rule uses 18 polar nodes per hemisphere, 40 radial nodes and 16 azimuthal nodes. The two solvers share the governing equations but no mesh, weak form, outer boundary or quadrature.

\section*{S2. Force and fold verification}

\paragraph{Spatial convergence.}
The half-domain and full-domain calculations use mirrored meshes and give pair and trimer forces within $0.01$ percent in eight tests. This checks the discrete symmetry conditions. Independent refinements in the 60-radius meridional box changed $\Delta F$ by at most $3\times10^{-5}$ in six cases as the surface element size went from 0.04 to 0.015. Table~\ref{tab:S4} shows two examples. Changing the mean-flow outer condition from zero mean velocity to stress-free changes the individual forces by under 0.3 percent at $D=3$ and 4 in these tests. At large spacing and high $\Omega$, the numerical spread can be comparable to $\Delta F$.

The force scan at $\Omega=3.8$ uses the reference 60-radius meridional box and mesh for the held chain and both isolated pairs at each of ten spacings between $D=4$ and 6.

\begin{table}[!htbp]
\centering
{\small
\begin{tabular}{cccccccc}
$\Omega$ & $D$ & element size & nodes (chain) & chain & pair($D$) & pair($2D$) & $\Delta F$ \\[3pt]
1 & 4 & 0.04 & 2608 & $-$0.29338 & $-$0.32631 & $-$0.02906 & +0.06199 \\
1 & 4 & 0.03 & 3491 & $-$0.29341 & $-$0.32635 & $-$0.02906 & +0.06200 \\
1 & 4 & 0.02 & 6479 & $-$0.29344 & $-$0.32638 & $-$0.02906 & +0.06201 \\
1 & 4 & 0.015 & 10631 & $-$0.29345 & $-$0.32639 & $-$0.02907 & +0.06201 \\
8 & 3 & 0.04 & 2485 & 0.05432 & 0.02712 & 0.00536 & +0.02184 \\
8 & 3 & 0.03 & 3343 & 0.05434 & 0.02713 & 0.00536 & +0.02185 \\
8 & 3 & 0.02 & 6119 & 0.05435 & 0.02714 & 0.00536 & +0.02185 \\
8 & 3 & 0.015 & 9994 & 0.05436 & 0.02714 & 0.00536 & +0.02186 \\
\end{tabular}}
\caption{Three-body increment $\Delta F=F_3-F_2(D)-F_2(2D)$ against the element size on the spheres (box 60).}
\label{tab:S4}
\end{table}

\begin{figure}[!htbp]
\centering
\includegraphics[width=\textwidth]{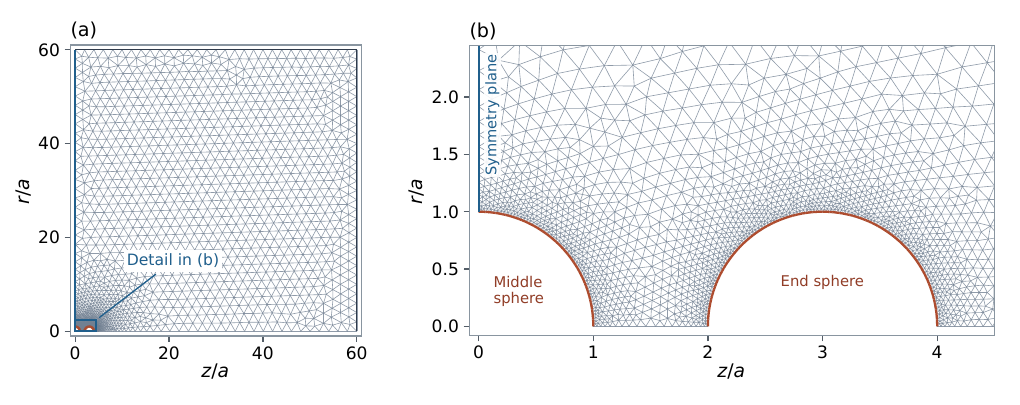}
\caption{Reference finite-element mesh for the held trimer in Table~\ref{tab:S4} at $D=3$ and sphere surface element size $0.03a$. (a) Meridional half-domain $0\le z/a\le60$, $0\le r/a\le60$, with reflection symmetry at $z=0$. (b) Local refinement at the middle sphere, the end sphere and the gap. The mesh has 3,343 vertices and 6,334 triangular elements.}
\label{fig:si_mesh}
\end{figure}

\paragraph{Finite-amplitude check.}
The time-domain and frequency-domain calculations share the mesh generator and element family. Using the same frequency-domain pair force at $2D$ in both increments, the five comparisons in Table~\ref{tab:S6} differ by 0.5 to 1.8 percent. The individual forces differ by up to 0.009 and the period means still change at the end of the runs. At $\Omega=3$, $D=3$, increasing the temporal resolution from 24 to 48 and 96 steps per period changes the pair offset from $-0.0213$ to $-0.0052$ and $-0.0012$. These runs check the last-period force increments at the sampled conditions. Estimating a finite-amplitude shift of a fold requires temporally converged force samples near that fold.

\begin{table}[!htbp]
\centering
{\small\setlength{\tabcolsep}{3pt}
\begin{tabular}{ccccccccccccc}
$\Omega$ & $D$ & $\epsilon$ & $h/a$ & box & \hd{steps/period\\$\times$ periods} & pair TD & pair FD & chain TD & chain FD & $\Delta F$ TD & $\Delta F$ FD & \hd{diff,\\\%} \\[3pt]
1 & 4 & 0.10 & 0.05 & 30 & $48\times8$ & $-$0.3354 & $-$0.3264 & $-$0.3020 & $-$0.2934 & 0.0625 & 0.0620 & 0.8 \\
3 & 3 & 0.10 & 0.05 & 30 & $48\times8$ & $-$0.2057 & $-$0.2006 & $-$0.1566 & $-$0.1521 & 0.0561 & 0.0555 & 1.1 \\
5 & 3.5 & 0.15 & 0.04 & 30 & $64\times16$ & $-$0.0071 & $-$0.0077 & 0.0132 & 0.0125 & 0.0205 & 0.0204 & 0.5 \\
8 & 2.5 & 0.15 & 0.04 & 30 & $64\times16$ & $-$0.0579 & $-$0.0580 & $-$0.0046 & $-$0.0054 & 0.0413 & 0.0405 & 1.8 \\
8 & 3 & 0.15 & 0.05 & 30 & $64\times16$ & 0.0285 & 0.0271 & 0.0558 & 0.0543 & 0.0220 & 0.0218 & 0.5 \\
\end{tabular}}
\caption{Finite-amplitude time-domain (TD) and frequency-domain (FD) forces in units of $\rho U^2a^2$, positive for attraction. Time-domain forces are averages over the last period. The excursion is $\epsilon=U/(\omega a)$, $h$ is the sphere surface element size and the box is given in radii. Both increments in the table use the frequency-domain pair force at $2D$.}
\label{tab:S6}
\end{table}

We also solved the far pair in the time domain at the same amplitude, mesh, time step and box. With all three forces from the time domain, the connected force differs from the frequency-domain value by $-2.2$ to $+0.7$ percent at the matched period count and by $-1.1$ to $+1.3$ percent when the final far-pair mean is used.

\paragraph{Folds of the force laws.}
We solve $F=\partial_D F=0$ for each held force law by iterating on a local polynomial of degree four in spacing and two in Stokes number. Table~\ref{tab:fold_mp} gives the resulting coordinates and derivatives. Changing the stencil half-widths from $(0.15,0.01)$ to $(0.1,0.02)$ and $(0.3,0.08)$ and the multipole order from 14 to 18 shifts $\Omega_c$ by at most $3\times10^{-7}$. For the pair and full chain, an independent finite-element procedure estimates the peak from a quadratic through $D=4.8$, 5.0 and 5.2 and interpolates the two peaks that bracket zero. Table~\ref{tab:M10} records the effects of mesh and box size on those thresholds. Using five spacings and quadratic, cubic or quartic fits gives a spread in peak forces that corresponds to about 0.002 in $\Omega$ at the local peak slope.

\begin{table}[!htbp]
\centering
{\small
\begin{tabular}{lrrrr}
Force law & $\Omega_c$ & $D_c$ & $10^3 A$ & $10^3 B$ \\[3pt]
Isolated pair & 4.082989 & 4.981590 & 7.71842 & 3.39218 \\
Pair-additive chain & 4.282943 & 4.802211 & 8.98879 & 4.42986 \\
Full chain & 3.533807 & 4.917458 & 10.67376 & 4.58429 \\
\end{tabular}}
\caption{Multipole folds of the held force laws. The local form is $F\simeq A(\Omega-\Omega_c)-B(D-D_c)^2$, with $A=\partial_\Omega F$ and $B=-\frac12\partial_D^2F$ evaluated at the fold.}
\label{tab:fold_mp}
\end{table}

\begin{table}[!htbp]
\centering
{\small
\begin{tabular}{cccc}
Force law & variant & $\Omega_c$ & shift against the base \\[3pt]
chain & $s_0=0.03$, box 60 & 3.53452 & +0.00000 \\
chain & $s_0=0.02$, box 60 & 3.53421 & $-$0.00031 \\
chain & $s_0=0.03$, box 120 & 3.53523 & +0.00072 \\
pair & $s_0=0.03$, box 60 & 4.08277 & +0.00000 \\
pair & $s_0=0.02$, box 60 & 4.08238 & $-$0.00038 \\
pair & $s_0=0.03$, box 120 & 4.08359 & +0.00082 \\
\end{tabular}}
\caption{Finite-element fold sensitivity to mesh and box size. We interpolate the force maxima between $\Omega=3.53$ and 3.54 for the full chain and between 4.08 and 4.10 for the pair.}
\label{tab:M10}
\end{table}

An unbounded point-singularity check of the additive fold uses the same reciprocal force identity with separate force samples and quadratures. Cubic and quartic fits give $\Omega_c$ from 4.28288 to 4.28293. The finest quartic gives $(\Omega_c,D_c)=(4.282929,4.802232)$ versus $(4.282943,4.802211)$ from multipoles. A finite-element stencil in the 60-radius box gives $(4.28063,4.82335)$, a difference of $0.00232$ in $\Omega$. At $(\Omega,D)=(3.8,9.6)$ the finite-element far-pair force exceeds the unbounded value by $2.1\times10^{-5}$. Dividing by the additive-fold slope in $\Omega$ gives a frequency scale of $0.0024$, consistent with sensitivity to the outer boundary.

\paragraph{Longitudinal modes.}
The finite-element calculations give the axial restoring stiffnesses in Table~\ref{tab:M11}. For breathing, we vary both spacings together and evaluate the slope of a local cubic fit to the attraction-positive outer-sphere force at its stable zero. For dimerisation, we increase one spacing and decrease the other by the same amount, using offsets $\delta a$ with $\delta=0.05$, 0.10 and 0.20. We fit the mean of the two end forces and the middle force along $z$ as odd cubics in $\delta$. The middle-sphere linear coefficient minus the end-sphere coefficient gives the restoring stiffness $k_a$. The additive comparison uses the two pair-force terms and its own stable zero. The full-chain breathing stiffness softens toward the fold.

\begin{table}[!htbp]
\centering
{\small
\begin{tabular}{ccccccc}
$\Omega$ & $D^*$ & $k_b$ & $k_a$ & $D^*$ additive & $k_b$ additive & $k_a$ additive \\[3pt]
3.6 & 4.5256 & 0.0052 & 0.0490 & none & none & none \\
3.8 & 4.1390 & 0.0146 & 0.0939 & none & none & none \\
4 & 3.9012 & 0.0236 & 0.1349 & none & none & none \\
4.5 & 3.5102 & 0.0472 & 0.2378 & 4.1586 & 0.0114 & 0.0348 \\
5 & 3.2513 & 0.0723 & 0.3422 & 3.6822 & 0.0321 & 0.1002 \\
6 & 2.9107 & 0.1233 & 0.5430 & 3.1946 & 0.0779 & 0.2474 \\
8 & 2.5253 & 0.2089 & 0.8380 & 2.7140 & 0.1717 & 0.5564 \\
\end{tabular}}
\caption{Finite-element held-chain axial stiffnesses at each force law's stable zero. $D^*$ is the spacing in sphere radii and $k_b$ and $k_a$ are the breathing and dimerisation stiffnesses in units of $\rho U^2a$.}
\label{tab:M11}
\end{table}

\section*{S3. Pair-force comparisons near the fold}

At the additive fold the pair spacings are about $4.80a$ and $9.60a$. We sample both under the same drive and boundary conditions as the trimer to resolve $F_2(D)+F_2(2D)=0$ and $F_2'(D)+2F_2'(2D)=0$. The prime denotes a spacing derivative at fixed $\Omega$. Forces below are in units of $\rho U^2a^2$, with attraction positive. The Fabre comparison uses held spheres and a reconstruction of the circular outer boundary with natural mean traction. The reciprocal and direct-traction forces are evaluated on the same reconstructed mesh.

\paragraph{Published force curves.}
At $D=2.5$, 3 and 4 and $\Omega=0.1$ to 100, our forces differ from Figure~6(a) of \citet{Fabre2017} by at most 0.02 and usually within the reading uncertainty of 0.01. At $\Omega=8$, the curve of \citet[Figure~3(b)]{ZhangRallabandi2025} gives about 0.0221 and 0.00527 at $D=3$ and 6, below our finite-element values of 0.02713 and 0.00536. Their Eq.~(6.5b) is a large-separation expansion through $D^{-4}$ for non-overlapping Stokes layers. At $\Omega=3.8$ the multipole force differs by 0.9 percent at $D=24$ and 0.09 percent at $D=48$, with a $D^{-5}$ difference. Near the fold spacings the estimated next term is comparable to the retained sum.

\paragraph{Fabre's zero at fixed spacing.}
At $D=4.48$, the vector trace of \citet[Figure~7]{Fabre2017} gives a zero at $\Omega=3.9999$. Our rectangular domain with zero mean velocity at its boundary gives 4.1998. A circular domain of their typical radius $40a$ with their natural mean traction condition gives 4.1252. The reconstruction reduces the fixed-spacing difference from 0.200 to 0.125. The original mesh and force samples are unavailable. At $(\Omega,D)=(4,4.48)$, the rectangular-domain force is $-2.620\times10^{-3}$. On the circular mesh the reciprocal integral and surface traction give $-1.670\times10^{-3}$ and $-1.673\times10^{-3}$. Zero mean velocity and the natural traction of the symmetric-strain form on that mesh give $-2.607\times10^{-3}$ and $-2.148\times10^{-3}$.

\paragraph{Folds of the pair force.}
Figure~7 of \citet{Fabre2017} traces the stable-spacing branch. The folds reported here solve $F_2=\partial_D F_2=0$ and can be compared across domains. In the circular reconstruction with natural mean traction, radii $40a$ and $80a$ give local folds at 3.94344 and 4.0667. Refining the $40a$ mesh from $0.03a$ to $0.02a$ gives 3.94319. Prescribing zero mean velocity at the $40a$ boundary gives 4.07949.

At the two low-frequency spacings from Fabre's Figure~7, $(\Omega,D)=(1,14.7998)$ and $(3,9.9319)$, the forces at radius $40a$ are $-1.93\times10^{-3}$ and $-1.29\times10^{-3}$. They remain repulsive at $80a$ and become attractive at $30a$, showing sensitivity to the far boundary at these spacings.

\section*{S4. Transverse stability and independent check}

\paragraph{Transverse derivatives.}
The derivative matrices are $K^x_{ij}=\partial F_{x,i}/\partial x_j$ and $K^y_{ij}=\partial F_{y,i}/\partial y_j$. Reflection gives the opposite displacement for each centred difference. Rotation about the drive axis makes $y$ tilt a null mode at equilibrium. For held spheres, the row sums are at most $2\times10^{-6}$ in the nine cases, and varying displacement, multipole order and azimuthal quadrature changes the eigenvalues by at most $1.1\times10^{-4}$ relative in the tests. The main text gives representative in-plane gradients. Table~\ref{tab:trans} compares normal bending across the multipole force gradients, the mobility check, the compressed-bond estimate and the additive chain at its own zero.

\begin{table}[!htbp]
\centering
{\small\setlength{\tabcolsep}{4pt}
\begin{tabular}{ccccccc}
$\Omega$ & $D^*$ & \hd{$y$ bend} & \hd{$y$ bend,\\Rotne-Prager} & \hd{$y$ bend,\\full mobility} & \hd{$-3\Phi(D^*)/D^*$} & \hd{$y$ bend,\\additive chain} \\[3pt]
3.6 & 4.5248 & +0.0047 & +0.0038 & +0.00375 & +0.0056 & none \\
4 & 3.9006 & +0.0107 & +0.0082 & +0.00801 & +0.0118 & none \\
4.5 & 3.5099 & +0.0179 & +0.0132 & +0.01288 & +0.0187 & $-$0.0010 \\
5 & 3.2511 & +0.0256 & +0.0182 & +0.01774 & +0.0255 & $-$0.00035 \\
8 & 2.5252 & +0.0745 & +0.0456 & +0.04350 & +0.0608 & +0.0093 \\
\end{tabular}}
\caption{Normal-bend force gradients of the held chain at the multipole equilibrium spacing $D^*$, in units of $\rho U^2a$. The two mobility columns give $y$-bend drift eigenvalues with steady mobility normalised by isolated-sphere drag. The full-mobility calculation includes the mean torque. The nearest-neighbour estimate uses the attraction-positive pair force $\Phi$. The additive comparison uses its own equilibrium, where one exists. Positive values denote instability.}
\label{tab:trans}
\end{table}

\begin{figure}[!htbp]
\centering
\includegraphics[width=0.79\textwidth]{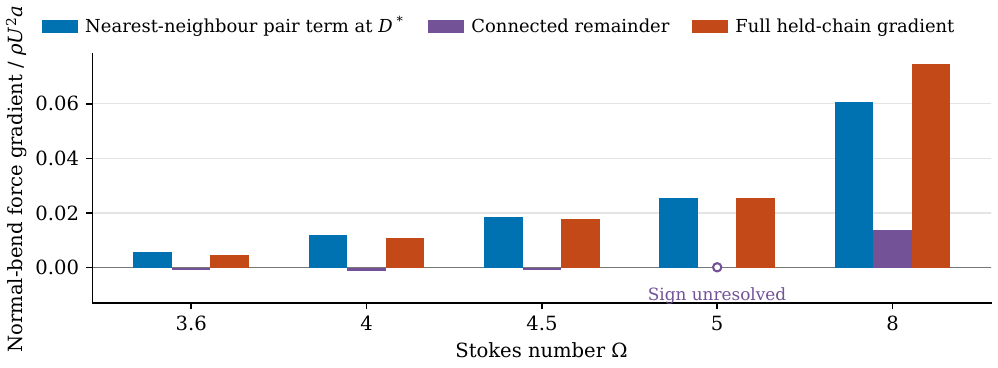}
\caption{Held normal-bend force gradient at the full-chain equilibrium. The pair contribution is evaluated at that same spacing. The connected remainder is the full gradient minus the pair contribution. Values are in units of $\rho U^2a$ and are calculated from the rounded entries in Table~\ref{tab:trans}. The open symbol at $\Omega=5$ marks a remainder whose sign is unresolved at this precision.}
\label{fig:compression}
\end{figure}

For the full-mobility column we evaluate the $18\times18$ steady Stokes resistance at the same equilibrium as the held load derivatives. The normal bend grows at all five spacings. Figure~\ref{fig:compression} shows the force-gradient decomposition at the full-chain spacing. The pair contribution ranges from 82 to 119 percent of the full gradient over these five equilibria. A comparison between the two equilibrium spacings is possible only where the additive law also has a restoring axial zero.

\paragraph{Mobility scales.}
Let $(\mathbf V_{\rm d},a\boldsymbol\omega_{\rm d})$ collect the dimensional slow translational and angular velocities of the spheres. We define the dimensionless resistance $\mathsf R^*$ by
\[
\begin{pmatrix}\mathbf F\\\mathbf T/a\end{pmatrix}
=\mu a\,\mathsf R^*
\begin{pmatrix}\mathbf V_{\rm d}\\a\boldsymbol\omega_{\rm d}\end{pmatrix},
\qquad
\widehat{\mathsf M}=6\pi(\mathsf R^*)^{-1},
\]
where $\mu=\rho\nu$ and $(\mathbf F,\mathbf T)$ are the loads needed to sustain the motion. The cycle-averaged streaming loads use scales $\rho U^2a^2$ for force and $\rho U^2a^3$ for torque. With those dimensionless loads denoted by $\mathbf F^*$ and $\mathbf T^*$, the normal drift matrix is $\widehat{\mathsf M}_{UF}\partial\mathbf F^*/\partial y^*+\widehat{\mathsf M}_{UT}\partial\mathbf T^*/\partial y^*$, where $y^*=y/a$ and the blocks select the normal translational velocity.

\paragraph{Free rotating chain.}
The disturbance on sphere $k$ satisfies $\mathbf V_k+\boldsymbol\Omega_k\times(\mathbf x-\mathbf x_{k0})-\mathbf e_x$, with the amplitudes set by Eq.~(3) of the main text. We extract the streaming loads as in Section S1 in the frame translating with the first-order motion of sphere $j$. The reciprocal forcing is $\mathbf b_j=\tfrac12\mathrm{Re}[(\overline{\mathbf v-\mathbf V_j}\cdot\nabla)\mathbf v]$. The mean surface correction on sphere $k$ is $-\tfrac12\mathrm{Re}[(\overline{\boldsymbol\xi_k-\boldsymbol\xi_j}\cdot\nabla)\mathbf v]$. The correction depends on centre displacement because rotation leaves the spherical boundary in place. Unit translations and rotations supply the auxiliary flows for the force and torque components. Their tractions also give the $6n\times6n$ Stokes resistance matrix at the on-axis equilibrium. We use the same mobility scales to convert derivatives of the streaming forces and torques into slow motion. Table~\ref{tab:S0b1} gives consistency checks. At $\Omega=4$ and 5 with $\tilde\rho=2.2$, varying the displacement from 0.01 to 0.04, the azimuthal cut-off, multipole order and quadrature changes the normal bending eigenvalue by at most $6.6\times10^{-4}$ relative. Table~\ref{tab:S0dModes} gives the coupled drift eigenvalues. At $\Omega=5$, density ratios 1.2, 1.5 and 8 give equilibrium spacings $D^*=2.96528$, 2.97583 and 3.17213, respectively. The normal-bend drift eigenvalues there are $6.63\times10^{-5}$, $3.67\times10^{-4}$ and $9.58\times10^{-3}$.

\begin{table}[!htbp]
\centering
{\small
\begin{tabulary}{\textwidth}{LL}
check & result \\[3pt]
free chain against a separate on-axis multipole implementation, $\Omega=5$, $\tilde\rho=2.2$, $D=3.2$ & outer-sphere force $6.31549815\times10^{-4}$ in both; volume and surface parts, $\mathbf V$ and the rocking rate to eight digits \\
held chain against the held off-axis code, $\Omega=5$, same $D^*$ & $K^x$, $K^y$ to $9\times10^{-14}$, torque derivatives to $5\times10^{-13}$ \\
straight chain: transverse forces and torques & at most $3.1\times10^{-15}$ and $1.8\times10^{-14}$ over 42 runs \\
translation null mode (row sums of $K^y$, $K^x$) & at most $7.8\times10^{-7}$ and $4.9\times10^{-6}$ \\
$y$-tilt null mode & at most $2.3\times10^{-5}$ absolute (entries up to 0.9) \\
$x$-tilt by a second route (drive tilted, chain undisplaced) & agrees to $5.5\times10^{-4}$ relative \\
free pair, sideways force against $-\Phi\,\delta y/D$ & $-2.3\times10^{-4}$, $-6.1\times10^{-5}$ ($\delta y=0.02$, 0.01; $\Omega=5$, $\tilde\rho=2.2$); $-1.3\times10^{-4}$, $-3.0\times10^{-5}$ ($\Omega=4$, $\tilde\rho=8$); $-2.6\times10^{-4}$, $-6.6\times10^{-5}$ (held, $\Omega=5$): the $O(\delta y^2)$ remainder \\
first-order boundary-condition residual & at most $4.5\times10^{-6}$ in the free-chain runs; $1.4\times10^{-5}$ and $1.4\times10^{-4}$ in two lower-order variants, which change the eigenvalue by $3\times10^{-7}$ and $7.5\times10^{-6}$ relative \\
\end{tabulary}}
\caption{Checks of the free off-axis route.}
\label{tab:S0b1}
\end{table}

\paragraph{Independent point-singularity calculation.}
Fundamental solutions centred on shells inside the spheres represent the oscillatory flow. Collocation enforces the moving boundary values. We integrate the pressure and viscous traction on each sphere to obtain the six rigid-body balances. The calculated complex translation of an isolated sphere is checked against the unsteady Stokes solution. Independent steady auxiliary fields give the mean streaming loads from the quadratic volume forcing and moving-surface slip. We partition the volume into Voronoi cells and integrate the far tail in a reciprocal radial coordinate.

We displace the centres by $h$ times the tilt or bend shape vector and project the load onto that vector, using reflection to obtain the opposite displacement. The reference spacings come from the multipole equilibria at $\tilde\rho=2.2$. We use $h=0.02a$ at $\Omega=3.6$ and $0.025a$ at $\Omega=8$. The axial quadrature clips each Voronoi cell on meridian arcs at fixed azimuth about the chain. The polar quadrature clips the azimuth intervals about the normal to the displacement plane. For the normal bend at both spacings and the in-plane modes at $\Omega=3.6$, the point-singularity entries in Table~\ref{tab:S0dModes} use the axial rule with 8 radial nodes per interval, 16 polar nodes per arc and 14 midpoint azimuths on the symmetry half-domain. The in-plane results at $\Omega=8$ use the polar rule with 7 radial, 6 polar and 14 azimuthal nodes per interval. We calculate the steady resistance from the point-singularity tractions and use it to include the mean torque in the normal bending drift.

At the unperturbed multipole spacings, refining the axial volume rule to $10\times20\times18$ and the traction quadrature gives end forces $+1.27\times10^{-7}$ at $\Omega=3.6$ and $-7.13\times10^{-6}$ at $\Omega=8$. Forces are scaled by $\rho U^2a^2$, with attraction positive. A denser source layout at $\Omega=8$ gives $+5.29\times10^{-6}$. Dividing the largest residual magnitudes by the local multipole force slopes gives spacing scales below $10^{-3}$ in $D$.

\begin{table}[!htbp]
\centering
{\small\setlength{\tabcolsep}{4pt}
\begin{tabular}{cclrrrrrr}
 & & & \multicolumn{3}{c}{Force gradient ($\times10^3$)} & \multicolumn{3}{c}{Drift eigenvalue ($\times10^3$)} \\
\cmidrule(lr){4-6}\cmidrule(lr){7-9}
$\Omega$ & $D^*$ & discretisation & $x$ tilt & $x$ bend & $y$ bend & $x$ tilt & $x$ bend & $y$ bend \\[3pt]
3.6 & 4.50346 & multipoles & $-1.0479$ & $-1.9306$ & $+0.2743$ & $-0.9365$ & $-1.5336$ & $+0.2176$ \\
 & & point singularities & $-1.0472$ & $-1.9291$ & $+0.2743$ & -- & -- & $+0.2174$ \\
4 & 3.69662 & multipoles & $-3.9023$ & $-7.7695$ & $+0.8463$ & $-3.3942$ & $-5.7831$ & $+0.6267$ \\
5 & 3.00643 & multipoles & $-12.7597$ & $-26.9525$ & $+2.3841$ & $-10.8090$ & $-18.3101$ & $+1.5889$ \\
6 & 2.67659 & multipoles & $-22.3551$ & $-48.1901$ & $+4.0307$ & $-18.6024$ & $-30.4862$ & $+2.4755$ \\
8 & 2.33541 & multipoles & $-39.1768$ & $-86.1270$ & $+7.0173$ & $-31.5823$ & $-48.1435$ & $+3.7882$ \\
 & & point singularities & $-38.7864$ & $-86.0531$ & $+6.9849$ & -- & -- & $+3.7704$ \\
\end{tabular}}
\caption{Transverse modes of freely translating and rotating chains at density ratio 2.2. Force gradients are scaled by $\rho U^2a$. Drift eigenvalues include mean torque and use steady mobility normalised by isolated-sphere drag. Negative values are restoring and positive values give growth. The point-singularity calculations use the multipole spacings and check normal drift at the two endpoint frequencies.}
\label{tab:S0dModes}
\end{table}

The tabulated results from the two discretisations agree within about 1 percent for the endpoint force gradients and normal drift (Table~\ref{tab:S0dModes}). At $\Omega=3.6$, doubling only $h$ changes the normal force gradient by 0.17 percent. Coarse polar runs at the same spacing used a different step and volume rule. They give a normal force gradient of $+0.000343$ and a torque-inclusive drift eigenvalue of $+0.000290$, about 25 and 34 percent above the multipole values. At $\Omega=8$, changing the volume quadrature from axial to polar at fixed step and surface settings changes the normal force gradient by 1.9 percent and the torque-inclusive drift eigenvalue by 4.3 percent. Normal bending grows in all these comparisons.

\paragraph{Growth time.}
Let $\lambda$ be a drift eigenvalue in Table~\ref{tab:S0dModes}, computed with the mobility scales above. Its dimensional growth rate is $\sigma=\lambda\rho U^2/(6\pi\mu)$. With drive period $T=2\pi/\omega$, this gives
\begin{equation}
\sigma T=\frac{\lambda\epsilon^2\Omega}{3},\qquad
N_e=\frac{1}{\sigma T}=\frac{3}{\lambda\epsilon^2\Omega}.
\label{eq:growth_time}
\end{equation}
For the free-chain normal bend at density ratio 2.2, the illustrative value $\epsilon=0.1$ gives $N_e=3.83\times10^5$ drive periods at $\Omega=3.6$ and $9.90\times10^3$ periods at $\Omega=8$. These are leading-order small-excursion estimates. The ordering also requires $\epsilon\ll D^*-2$; at $\Omega=8$, the gap is only 0.3354 sphere radii. Reducing $\epsilon$ from 0.1 to 0.03 increases the growth time by a factor of 11.1, giving $N_e=1.10\times10^5$ periods at this point.

\FloatBarrier
\bibliographystyle{IEEEtranN}
\bibliography{supplement_references}